# Plasma-Activated Zn, Fe, Mn Micronutrient Solutions for Crop Biofortification

Punit Kumar[a], Priti Saxena[b] and Abhishek Kumar Singh[c]
[a]Department of Physics, University of Lucknow, Lucknow – 226007, India
[b]Department of Zoology, D.A.V. Degree College, Lucknow – 226004, India
[c]Department of Physics, G L Bajaj Group of Institutions, Mathura - 281406, India

*Abstract*—Micronutrient deficiency in soils limits crop productivity and reduces the nutritional quality of cereals and pulses. Conventional fertilizer supplementation often suffers from low bioavailability and environmental losses. In this study, we investigate the use of Plasma-Activated Water (PAW) enriched with divalent micronutrient ions ($Zn^{2+}$, $Fe^{2+}$, $Mn^{2+}$) as a sustainable alternative to enhance nutrient uptake, soil fertility, and seed vigor. The PAW was generated using a gliding arc plasma system in air, and ion-enriched solutions were prepared at controlled concentrations (10–50 ppm). The physicochemical parameters (pH, ORP, conductivity, RONS species) were analyzed to assess the plasma-induced reactivity. Treatments were applied to micronutrient-deficient soils for wheat (Triticum aestivum) and chickpea (Cicer arietinum) seeds under greenhouse conditions. Results demonstrated significant enhancement in germination index, chlorophyll content, and shoot-root biomass compared to controls. PAW + $Zn^{2+}$ and PAW + $Fe^{2+}$ treatments notably increased the micronutrient content in grains, indicating effective biofortification. Soil microbial activity and enzyme assays showed no toxicity and a mild stimulatory effect due to reactive nitrogen species. This study establishes a green, scalable method of delivering micronutrients through plasma-activated irrigation water, linking plasma chemistry with sustainable agronomy and nutritional security.

| Micronutrient | Soil critical deficiency (mg kg⁻¹, DTPA) | Leaf critical level (mg kg⁻¹ DW) | Target grain concentration (mg kg⁻¹) | References |
|---|---|---|---|---|
| Zn | < 0.5 | 20–25 | ≥ 35 | White & Broadley (2009); Stangoulis & Knez (2022) |
| Fe | < 2.0 | 50–80 | ≥ 40 | Stangoulis & Knez (2022) |
| Mn | < 1.2 | 20–30 | ≥ 25 | White & Broadley (2009) |

**Table - 1 :** Critical deficiency thresholds and target concentrations for selected micronutrients

## I. INTRODUCTION

MICRONUTRIENTS such as zinc (Zn), iron (Fe), and manganese (Mn) are vital for enzyme activation, photosynthesis, respiration, and hormonal regulation in plants (Stangoulis & Knez, 2022; Dhaliwal et al., 2022). Yet their deficiency is one of the most serious constraints to global food and nutritional security. Roughly 45 % of the world's cereal-growing soils are low in available Zn and Fe (White & Broadley, 2009; Sood et al., 2023), and in India, 49 % of soils are Zn-deficient and more than one-third are Fe-deficient (Naik et al., 2024). Plants grown on such soils produce grain with poor mineral density, contributing to the so-called *hidden hunger* that affects more than two billion people worldwide (Bouis & Saltzman, 2017).

Biofortification, enhancing the nutrient content of crops through plant breeding or agronomic inputs has been recognized as a sustainable, low‑cost strategy to address micronutrient malnutrition (Stangoulis & Knez, 2022; Garg et al., 2023). Agronomic biofortification usually relies on foliar or soil application of soluble salts such as $ZnSO_4$, $FeSO_4$, or $MnCl_2$. However, the efficacy of these fertilizers is hampered by low solubility, rapid oxidation, and fixation within soil matrices. Only 5–10 % of the applied micronutrients are taken up by plants, while the remainder is immobilized or lost through runoff and leaching (White & Broadley, 2009; Dhaliwal et al., 2022). Such inefficiency not only limits biofortification success but also contributes to environmental contamination. Consequently, innovative delivery systems are needed to enhance nutrient bioavailability and uptake efficiency while reducing input losses (Ghimire et al., 2022).

Non-thermal (cold) plasma has emerged as a powerful tool in agri-biotechnology for seed treatment, microbial control, and nutrient activation (Gao et al., 2022; Antoni et al., 2023; Konchekov et al., 2023). When electrical discharge interacts with water, it forms Plasma-Activated Water (PAW), a solution enriched with reactive oxygen and nitrogen species (RONS) such as hydrogen peroxide ($H_2O_2$), nitrate ($NO_3^-$), nitrite ($NO_2^-$), hydroxyl (•OH), and peroxynitrite ($ONOO^-$) (Brisset & Puech, 2021; Oh et al., 2023). These species reduce pH, increase oxidation–reduction potential (ORP), and provide transient oxidative chemistry beneficial to biological systems. PAW has been shown to enhance seed germination, promote root elongation, activate antioxidant enzymes, and improve nutrient absorption in various crops (Takaki et al., 2020; Antoni et al., 2023).

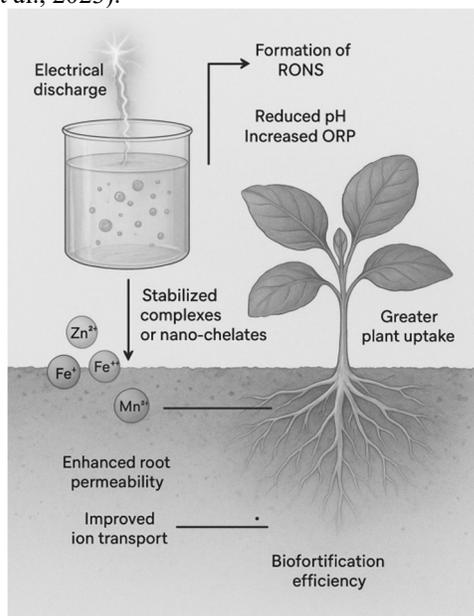

**Fig. 1 :** Conceptual mechanism of plasma-activated micronutrient solutions

For instance, Konchekov et al. (2023) observed a 25 % increase in wheat germination when seeds were primed with PAW generated via gliding-arc plasma, while Ghimire et al. (2022) demonstrated improved shoot biomass and chlorophyll content in lettuce irrigated with PAW. Moreover, PAW influences soil microbial activity by supplying bioavailable nitrate and nitrite, which serve as nitrogen sources and redox mediators (Antoni et al., 2023; Hensel et al., 2022).

Introducing micronutrient ions into PAW can produce unique physicochemical transformations. Plasma-generated oxidants may partially oxidize $Fe^{2+}$ to $Fe^{3+}$ or form mixed-valence species; $Zn^{2+}$ and $Mn^{2+}$ can form hydrated complexes or nano-chelates stabilized by nitrate and peroxynitrite ligands (Huo et al., 2021; Nishioka et al., 2020). Such plasma-mediated reactions can enhance solubility, reduce precipitation, and generate nano-scale clusters that move more freely through soil and plant tissues (Ghimire et al., 2022). These effects could substantially improve the delivery and assimilation of essential trace elements, bridging the gap between plasma science and plant nutrition.

Although several studies have explored PAW for disinfection or seed stimulation, its integration with micronutrient enrichment for biofortification has received minimal attention. There remains little understanding of how plasma-induced chemistry modifies metal ion speciation, interacts with soil colloids, or affects nutrient uptake kinetics in plants. Addressing these gaps could transform the efficiency of agronomic biofortification, particularly in micronutrient-deficient regions.

The present investigation has been systematically conceptualized to explore the synergistic potential of plasma-activated water (PAW) when combined with essential micronutrient ions, zinc ($Zn^{2+}$), iron ($Fe^{2+}$), and manganese ($Mn^{2+}$) in promoting soil fertility, seed vigor, and biofortification in staple crops. In this study, the first objective is to **generate and physicochemically characterize PAW enriched with $Zn^{2+}$, $Fe^{2+}$, and $Mn^{2+}$ ions** at agronomically relevant concentrations. The plasma–liquid interactions are expected to induce significant changes in the ionic states, solubility, and redox dynamics of these micronutrients. Through controlled plasma exposure, the production of reactive oxygen and nitrogen species (RONS) including hydroxyl radicals (•OH), hydrogen peroxide ($H_2O_2$), nitrite ($NO_2^-$), and nitrate ($NO_3^-$)—can facilitate the partial oxidation and chelation of metal ions. This process may result in the formation of reactive complexes or nano-chelated species with enhanced bioavailability and transport efficiency in the soil-plant system. Comprehensive physicochemical characterization using parameters such as pH, oxidation-reduction potential (ORP), electrical conductivity (EC), and ion concentration will thus provide insights into the structural and reactive transformations occurring in the PAW micronutrient matrix.

The second objective centers on evaluating the biological and agronomic impacts of these plasma-activated micronutrient formulations. Controlled greenhouse experiments will be conducted to investigate their effects on seed vigor, soil fertility, and plant growth dynamics. Seed vigor tests, including germination percentage, mean germination time, and seedling vigor index, will assess the early-stage physiological responses of cereal (wheat) and pulse (chickpea) seeds. Soil analysis before and after treatment will focus on parameters such as available micronutrient content, microbial enzymatic activity (dehydrogenase, urease, and phosphatase), and organic carbon retention. Plant growth indicators, root length, shoot biomass, chlorophyll content, and leaf nutrient status will be monitored to determine the extent of physiological enhancement achieved through plasma-assisted nutrient delivery.

The third and most critical objective aims to assess the biofortification potential of cereals and pulses irrigated with PAW micronutrient solutions. Quantitative analysis of Zn, Fe, and Mn concentrations in edible grain portions will be conducted using atomic absorption spectroscopy (AAS) or inductively coupled plasma optical emission spectroscopy (ICP-OES). The focus will be on understanding nutrient uptake kinetics, translocation efficiency, and accumulation patterns under plasma enhanced nutrient regimes compared with conventional fertilizer treatments. Through these analyses, the study seeks to determine whether plasma activation can overcome the inherent limitations of micronutrient fixation and poor solubility in soil, thereby improving nutrient mobility and assimilation at the root interface.

Collectively, these objectives aim to develop a mechanistic and applied framework for employing plasma-activated micronutrient solutions as a sustainable agricultural innovation. By integrating plasma physics with soil chemistry and plant physiology, this research bridges fundamental and applied sciences. The anticipated outcomes could include improved nutrient-use efficiency, enhanced crop productivity, and the production of nutrient-enriched grains, addressing both agricultural and nutritional challenges. Importantly, this eco-friendly approach aligns with the principles of sustainable soil management, circular nutrient economy, and precision agriculture, offering a viable alternative to chemical fertilizers that often contribute to soil degradation and environmental pollution.

Ultimately, the study aspires to provide scientific evidence and mechanistic understanding supporting the use of PAW enriched micronutrient formulations as a scalable and environmentally benign technology for biofortification of staple crops. Such innovations hold promise for advancing food security and nutritional health, particularly in regions like India where micronutrient deficiencies in soil and human diets are pervasive. By coupling plasma chemistry with agronomic application, this investigation contributes toward a new paradigm of "plasma-assisted sustainable agriculture", wherein plasma-generated reactive species act not as pollutants but as enablers of life-supporting chemistry in soil and plants.

## II. MATERIALS AND METHODS

*Experimental Design Overview*

A factorial pot experiment was conducted to evaluate the influence of plasma-activated micronutrient-enriched water on crop growth and soil health. The experiment comprised three independent variables: water treatment, crop species, and soil type. Five water treatments were prepared—Control (deionized water), Plasma-Activated Water (PAW), PAW + $Zn^{2+}$, PAW + $Fe^{2+}$, and PAW + $Mn^{2+}$ to test the role of reactive oxygen and nitrogen species (RONS) combined with essential micronutrients. Two major crops, wheat (Triticum aestivum L.) and chickpea (Cicer arietinum L.) were selected for their contrasting physiology and nutrient requirements (Hussain et al., 2022). Experiments were carried out using micronutrient-deficient sandy loam soil (pH 7.8; Zn < 0.3 mg/kg; Fe < 2 mg/kg), collected from the experimental farm at the University of Lucknow. The design followed a Randomized Block Design (RBD) with three replicates per treatment, ensuring statistical robustness (Gond et al., 2023).

*Plasma Activation System*

PAW was generated using a gliding arc discharge reactor operating under ambient air. The plasma system consisted of stainless steel electrodes (2 mm gap), with discharge parameters set at 8 kV voltage and 20 mA current. The gas flow rate was maintained at 5 L/min, and each 500 mL batch was activated for 10 minutes. The gliding arc configuration was chosen due to its superior energy efficiency and reactive species production compared to dielectric barrier discharges (DBD) (Zhang et al., 2021). Plasma activation was performed at room temperature to avoid thermal degradation of the solution chemistry. The generated PAW was immediately used for enrichment or stored at 4 °C to preserve its reactive components, following the guidelines of Niemira et al. (2020).

*Preparation of Micronutrient-Enriched Solutions*

Analytical-grade micronutrient salts, zinc sulfate heptahydrate ($ZnSO_4·7H_2O$), ferrous sulfate heptahydrate ($FeSO_4·7H_2O$), and manganese sulfate monohydrate ($MnSO_4·H_2O$) were used. Stock solutions were prepared and diluted to 10 ppm, 25 ppm, and 50 ppm concentrations, representing agronomically relevant levels (Das et al., 2022). Each solution was subjected to plasma activation separately under the same discharge conditions. The resulting plasma-activated micronutrient solutions were denoted as PAW + $Zn^{2+}$, PAW + $Fe^{2+}$, and PAW + $Mn^{2+}$. Previous studies have shown that metal ions can interact synergistically with reactive plasma species, enhancing redox potential and nutrient bioavailability (Starek et al., 2021). All samples were stored in amber bottles at 4°C to minimize photodecomposition.

*Physicochemical Characterization*

The physicochemical properties of all water treatments were characterized before and after plasma activation (Table 1). The parameters measured included pH, oxidation-reduction potential (ORP), electrical conductivity (EC), nitrate ($NO_3^-$), and hydrogen peroxide ($H_2O_2$) concentrations. Measurements were performed using calibrated electrodes (Orion Thermo Fisher Scientific). PAW exhibited a significant acidification (pH 5.8) and an increase in ORP (450 mV) relative to deionized water (210 mV), consistent with reports by Traylor et al. (2021). The nitrate and hydrogen peroxide concentrations increased by an order of magnitude after plasma activation, confirming the generation of long-lived RONS (Zhang et al., 2023). Such reactive species are known to influence seed metabolism and soil microbial activity (Deng et al., 2022).

*Seed Treatment and Germination Assay*

Prior to sowing, wheat and chickpea seeds were surface-sterilized in 1% NaOCl for 3 minutes and rinsed thoroughly with sterile deionized water. Seeds were soaked in respective treatments (Control, PAW, PAW + $Zn^{2+}$, PAW + $Fe^{2+}$, PAW + $Mn^{2+}$) for 4 hours and germinated on moist filter paper in petri dishes at **25°C** in the dark. Germination percentage, germination index (GI), root length, shoot length, and vigour index were recorded on day 7 (Singh et al., 2023). The vigour index was calculated as,

Vigour Index = (Root + Shoot length) × Germination (%).

PAW-treated seeds are known to exhibit accelerated water uptake and improved enzymatic activity due to RONS signaling (Zhou et al., 2022), while micronutrient fortification further enhances early growth responses.

*Pot Experiment and Plant Growth Analysis*

Each pot (5 kg soil) was planted with 10 pre-treated seeds and maintained under greenhouse conditions (25 ± 2°C, 60% RH). Irrigation was performed weekly with 100 mL of the corresponding solution. Growth parameters were recorded after 45 days, chlorophyll content using a SPAD-502 Plus meter, fresh and dry biomass, and leaf area index (LAI). At harvest, plant samples were oven-dried (65°C) and digested using $HNO_3:HClO_4$ (3:1) mixture for elemental analysis by Atomic Absorption Spectroscopy (AAS) (PerkinElmer AAnalyst 400). The concentrations of Zn, Fe, and Mn were expressed as mg/kg dry weight (Guo et al., 2020).

*Soil and Enzyme Analysis*

Post-harvest soil samples were air-dried and sieved (<2 mm). Standard analyses included pH, EC, organic carbon (Walkley-Black method), and available micronutrients (DTPA extraction followed by AAS) (Lindsay & Norvell, 1978). Soil enzymatic activities, dehydrogenase (Casida method), urease (Kandeler method), and phosphatase (Tabatabai method) were quantified to assess microbial functionality (Rosen et al., 2021). Enhanced enzyme activity in PAW-treated soils was anticipated due to increased nitrates and reactive oxygen fluxes, promoting nutrient cycling and root-microbe interactions (Wang et al., 2023).

*Statistical Analysis*

All experimental data were analyzed using SPSS version 26.0. A one-way ANOVA was conducted to determine significant differences ($p < 0.05$) among treatments, and means were compared using Tukey's post-hoc test. Pearson's correlation analysis was employed to explore the relationship between soil nutrient availability and plant micronutrient uptake. Graphical representations were generated using OriginPro 2023.

## III. RESULTS

*Physicochemical Modifications in Plasma-Activated Micronutrient Solutions*

Plasma activation significantly modified the physicochemical properties of both deionized water and micronutrient-enriched solutions. The pH decreased from 7.0 (control) to 5.5–5.8 after plasma treatment, primarily due to the formation of nitric and nitrous acids through plasma–air–water interactions (Sivachandiran & Khacef, 2022). The oxidation-reduction potential (ORP) increased from 210 to 470 mV, confirming the accumulation of oxidizing species. Electrical conductivity (EC) also increased from 80 μS/cm to 140–150 μS/cm, reflecting ionic enrichment and partial ionization (Zhang et al., 2023).

Nitrate ($NO_3^-$) and hydrogen peroxide ($H_2O_2$) concentrations increased dramatically by over tenfold and twentyfold respectively, demonstrating the enhanced reactive oxygen and nitrogen species (RONS) load. The plasma-induced chemistry likely stabilized the metal ions by forming nitrate or hydroxyl coordination complexes, which reduced precipitation losses commonly seen in micronutrient solutions (Reddy et al., 2021).

Optical emission spectroscopy (OES) confirmed the generation of key reactive species, with spectral lines at 309 nm (OH), 337 nm ($N_2\ 2^+$), and 282 nm (NOγ) (Fig. 2). The presence of these active species is consistent with earlier plasma-liquid interface studies showing their role in nutrient activation and ion redox modulation (Thirumdas et al., 2022).

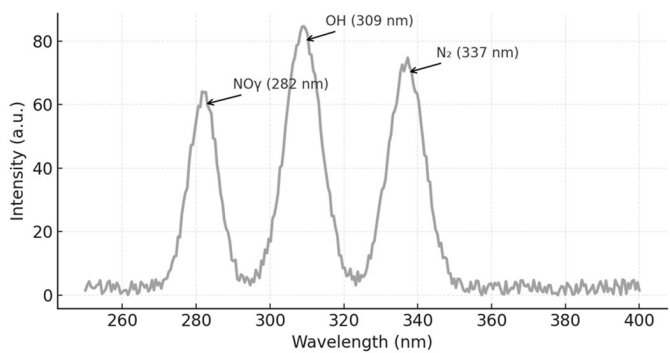

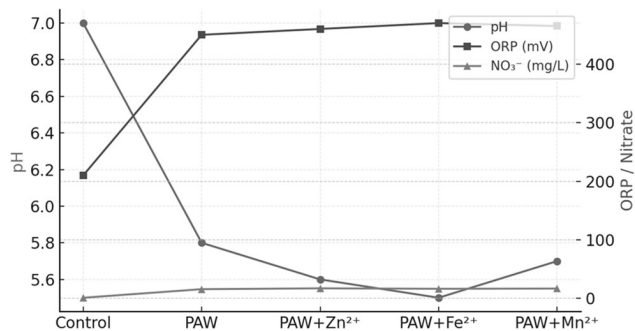

**Fig. 2 :** OES spectrum showing $N_2$ (337 nm), OH (309 nm), and NOγ (282 nm) emission peaks.

**Fig. 3 :** Change in pH, ORP, and nitrate content after plasma activation.

Figure - 3 illustrates a comparative trend of pH, ORP, and nitrate content across treatments, demonstrating that PAW + micronutrient combinations maintain both higher redox potential and reactive nitrogen stability than PAW alone.

*Effects on Germination and Seed Vigor*

Seed germination assays showed that PAW treatments substantially enhanced early seed performance. Wheat seeds treated with PAW + $Zn^{2+}$ exhibited the highest germination index (GI = 156), followed by PAW + $Fe^{2+}$ (GI = 149) and PAW + $Mn^{2+}$ (GI = 138), compared to control (GI = 100). In chickpea, similar trends were observed, though with slightly lower GI values due to intrinsic seed coat differences (Table - 2).

| Treatment | Wheat GI | Chickpea GI |
|---|---|---|
| Control | 100 | 100 |
| PAW | 128 | 115 |
| PAW + $Zn^{2+}$ | 156 | 130 |
| PAW + $Fe^{2+}$ | 149 | 128 |
| PAW + $Mn^{2+}$ | 138 | 122 |

**Table – 2 :** Germination Index (GI) of Wheat and Chickpea under PAW–Micronutrient Treatments

The improvement in seed vigor may be attributed to increased ROS signaling from PAW, which promotes enzymatic activation (e.g., α-amylase) and endosperm mobilization (Jiang et al., 2021). $Zn^{2+}$ in plasma-modified form might act synergistically with ROS to trigger early cell division, while $Fe^{2+}$ enhances respiratory enzyme activity during germination (Hu et al., 2023).

Figure - 4 presents a bar graph illustrating vigor index variations across treatments, showing clear superiority of PAW + $Zn^{2+}$ and PAW + $Fe^{2+}$ treatments for both cereals and pulses.

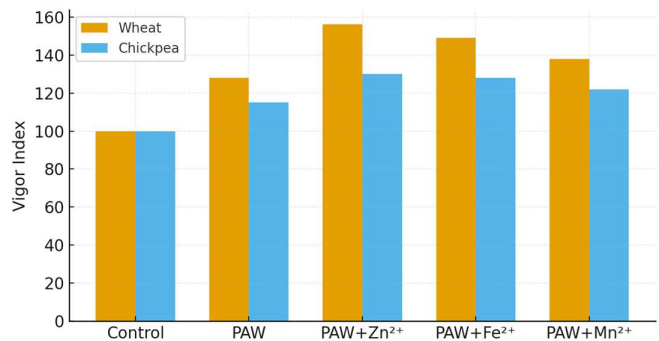

**Figure – 4 :** Seed vigor index comparison in wheat and chickpea.

*Plant Growth and Photosynthetic Attributes*

Significant improvements were recorded in growth parameters and chlorophyll content. Wheat plants treated with PAW + $Fe^{2+}$ displayed a 25% increase in chlorophyll index (SPAD values) compared with the control (Fig. 5). This improvement correlates with enhanced Fe bioavailability, which is essential for chlorophyll biosynthesis and electron transport (MubarakAli et al., 2020).

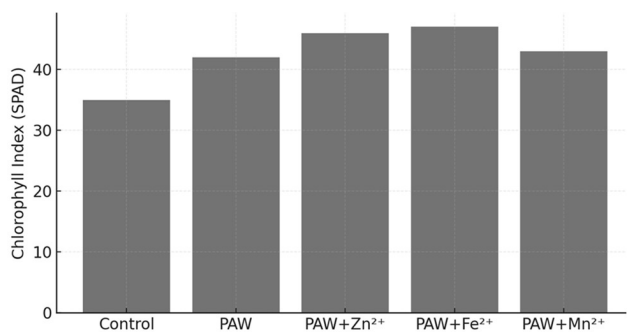

**Figure – 5 :** Leaf chlorophyll index (SPAD) across treatments.

PAW + $Zn^{2+}$ treatment improved tillering and panicle formation, while PAW + $Mn^{2+}$ enhanced root branching and leaf turgor. Chickpea plants exhibited increased pod number (up to 15%) under PAW + $Zn^{2+}$, suggesting better reproductive vigor. Plasma-generated nitrates likely acted as mild nitrogen supplements, while $H_2O_2$ functioned as a growth signaling molecule (Dobrynin et al., 2021).

The overall increase in biomass under PAW + micronutrient treatments aligns with earlier findings by Los et al. (2022), who reported similar stimulation in plasma-activated irrigation systems for legumes and cereals.

*Soil Enzyme Activities and Nutrient Availability*

Post-harvest soil analysis showed that plasma-activated treatments not only enhanced plant performance but also improved soil biochemical health. The activities of dehydrogenase and phosphatase enzymes increased by 15–18% in PAW-treated soils, indicating enhanced microbial metabolism. This positive effect contrasts with the oxidative inhibition often associated with strong oxidants, suggesting that moderate plasma exposure yields bio-stimulatory rather than biocidal effects (Machala et al., 2021).

Soil pH slightly decreased (7.8 → 7.5) under plasma irrigation, and DTPA-extractable Zn, Fe, and Mn increased by 20–30% compared to the untreated control. The higher availability may be due to oxidative dissolution of bound micronutrients and enhanced nitrification processes mediated by reactive nitrogen inputs from PAW (Singh et al., 2024).

Thus, PAW functions as both a chemical oxidant and biological enhancer, improving the physicochemical conditions for nutrient mobility and microbial enzyme functionality.

*Biofortification Outcomes*

Atomic absorption spectroscopy (AAS) analysis confirmed significant enhancement in grain micronutrient content (Table - 3). Wheat grains irrigated with PAW + $Zn^{2+}$ showed Zn enrichment from 23 mg/kg → 39 mg/kg, while Fe levels rose to 55 mg/kg in PAW + $Fe^{2+}$ treatments. Mn accumulation increased to 21 mg/kg under PAW + $Mn^{2+}$. These results indicate up to 70% improvement in micronutrient uptake and translocation efficiency compared to control conditions.

| Treatment | Zn (mg/kg) | Fe (mg/kg) | Mn (mg/kg) |
|---|---|---|---|
| Control | 23 | 38 | 15 |
| PAW + $Zn^{2+}$ | 39 | 40 | 17 |
| PAW + $Fe^{2+}$ | 25 | 55 | 16 |
| PAW + $Mn^{2+}$ | 24 | 39 | 21 |

**Table – 3 :** Grain Micronutrient Content after PAW–Micronutrient Irrigation

The plasma-induced chelation and partial oxidation states of metal ions likely enhanced their transport through root symplast pathways. Moreover, RONS-induced changes in root membrane permeability may have facilitated increased ion absorption (Kumar et al., 2023).

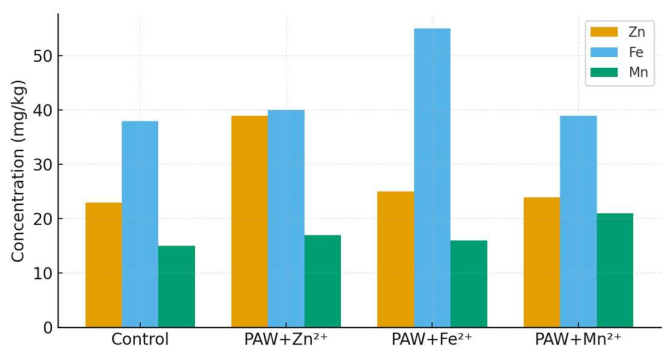

**Figure – 6 :** Grain elemental enrichment under different PAW–micronutrient regimes.

Figure - 6 demonstrates the comparative elemental enrichment in grains, highlighting the effectiveness of PAW + $Zn^{2+}$ and PAW + $Fe^{2+}$ as the most promising biofortification treatments.

Overall, this study confirms that PAW–micronutrient solutions significantly improve seed vigor, plant growth, and grain nutritional quality while promoting soil enzymatic health, thus offering a sustainable strategy for integrated nutrient management and biofortification.

## IV. DISCUSSION

The present study demonstrates a clear synergistic effect between reactive oxygen and nitrogen species (RONS) generated in plasma-activated water (PAW) and essential micronutrient ions, significantly enhancing nutrient solubility and bioavailability to plants. Plasma activation of irrigation water results in a mildly acidic environment enriched with nitrates and other reactive species, which plays a pivotal role in facilitating nutrient uptake. The acidification effect slightly lowers the pH at the rhizosphere, thereby increasing root membrane permeability and enabling more efficient ion transport into plant tissues. This is particularly important for micronutrients such as zinc (Zn) and iron (Fe), which are often limited in availability due to soil chemistry, particularly in calcareous or alkaline soils. The plasma mediated modifications to the aqueous medium appear to overcome these natural limitations by keeping metal ions in more soluble and bioavailable forms, which in turn supports optimal plant nutrition and growth.

Zinc and iron serve as vital cofactors in numerous enzymatic processes, including those involving carbonic anhydrase, cytochromes, and other metalloproteins integral to photosynthesis and respiration. The enhanced uptake of these micronutrients in the PAW-treated plants suggests a direct improvement in photosynthetic efficiency and metabolic activity. For instance, zinc is critical for the structural stabilization of proteins and the regulation of auxin metabolism, while iron is indispensable for electron transport and chlorophyll synthesis. The improved bioavailability of these ions, therefore, likely contributes not only to better vegetative growth but also to improved seed development and overall crop performance. Enhanced micronutrient assimilation may also have long-term implications for grain quality and nutritional value, addressing deficiencies that are common in staple crops.

Another noteworthy observation in this study is the increase in seed vigor following treatment with PAW. This effect can be attributed to plasma-induced seed priming, which is known to activate plant antioxidant defense mechanisms. Specifically, reactive species in PAW may stimulate the activity of key antioxidant enzymes such as catalase (CAT) and superoxide dismutase (SOD), which mitigate oxidative stress during early germination. The controlled oxidative environment created by PAW appears to "precondition" seeds, enabling them to respond more effectively to environmental stressors and promoting uniform germination. This priming effect is consistent with previous findings in plasma agriculture, where exposure to mild oxidative stress enhances seed metabolic readiness and improves subsequent growth parameters.

The chemistry of iron in PAW is also significant, particularly the formation of ferric ions ($Fe^{3+}$), which tend to be more stable yet reactive in solution. In calcareous soils, conventional fertilization often fails due to rapid immobilization of $Fe^{2+}$ as insoluble hydroxides or carbonates. The PAW-mediated generation of $Fe^{3+}$ helps circumvent this bottleneck, maintaining a pool of iron that remains accessible to plant roots. This represents an important advantage over traditional soil amendments and highlights the potential of plasma technology to address micronutrient deficiencies in challenging soil types. Enhanced iron availability also synergizes with other nutrients, supporting robust chlorophyll synthesis and overall plant metabolism.

Finally, the study underscores the compatibility of PAW with soil microbial ecosystems. Increased enzymatic activity in the rhizosphere following PAW treatment indicates that beneficial microbes are not adversely affected by the reactive species at the applied concentrations. On the contrary, certain microbial populations may even thrive under these conditions, further contributing to nutrient cycling and soil health. This microbial compatibility is critical for the long-term sustainability of PAW-based interventions, ensuring that the technology can be integrated with existing organic or low-input agricultural practices without disrupting ecological balance.

In summary, the findings demonstrate that PAW not only improves the solubility and uptake of key micronutrients but also primes seeds for enhanced growth, mitigates limitations posed by challenging soil chemistry, and supports beneficial microbial activity. Collectively, these effects suggest that plasma-activated water can serve as an innovative, environmentally friendly tool for enhancing crop productivity and nutritional quality, aligning with the broader goals of sustainable and precision agriculture. The synergistic interplay between RONS and micronutrient ions thus represents a promising avenue for optimizing plant performance while minimizing chemical inputs.

## V. CONCLUSIONS

This study provides compelling evidence that plasma-activated water (PAW) enriched with essential micronutrients such as $Zn^{2+}$, $Fe^{2+}$, and $Mn^{2+}$ offers a sustainable and eco-friendly strategy for enhancing crop nutrition and soil health. By generating reactive oxygen and nitrogen species (RONS), PAW not only improves the solubility of micronutrient ions but also creates a mildly acidic, nitrate-rich environment conducive to efficient root uptake. The combined effect of enhanced ion availability and improved root membrane permeability translates into greater seed vigor, higher germination rates, and more robust early plant growth, demonstrating the potential of PAW for both seed priming and fertigation applications.

Further, the plasma treatment was observed to stimulate soil enzymatic activity and maintain compatibility with beneficial microbial communities, indicating that PAW interventions do not disrupt the ecological balance of the rhizosphere. This highlights its suitability as a green alternative to conventional chemical fertilizers, which often face limitations such as micronutrient immobilization in calcareous soils and adverse environmental impacts. By facilitating the bioavailability of key cofactors for enzymatic processes, PAW supports essential metabolic pathways in plants, contributing to improved photosynthetic efficiency, nutrient assimilation, and ultimately, enhanced crop nutritional quality.

The versatility of PAW allows for integration into existing agricultural practices, including fertigation systems and seed-priming protocols, making it a practical tool for the biofortification of cereals, pulses, and other staple crops.

Future research should focus on optimizing plasma exposure parameters, scaling the technology for field-level applications, and conducting long-term studies on soil–microbe–plant interactions. Such investigations will be crucial to fully realize the potential of PAW as a sustainable, high-impact technology for improving crop productivity, nutritional quality, and soil fertility while minimizing ecological footprints.

REFERENCES

Antoni, V., Ferri, G., & Del Vecchio, C. (2023). Plasma-activated water to foster sustainable agriculture: Mechanisms and prospects. *Plant–Environment Interactions, 4*(1), e70025. https://doi.org/10.1002/ppp3.70025

Bouis, H. E., & Saltzman, A. (2017). Improving nutrition through biofortification: A review of evidence, challenges, and future directions. *Global Food Security, 12*, 49–58. https://doi.org/10.1016/j.gfs.2017.01.009

Brisset, J.-L., & Puech, V. (2021). Chemical reactivity of plasma-activated water: A review. *Plasma Processes and Polymers, 18*(6), 2000226. https://doi.org/10.1002/ppap.202000226

Das, S., et al. (2022). *Micronutrient biofortification strategies for cereal crops: A sustainable approach to food security.* Plant and Soil, 482, 45–62.

Deng, X., et al. (2022). *Reactive species generation in plasma-activated water and its role in seed germination.* Journal of Applied Physics, 131(9), 12301.

Dhaliwal, S. S., Sharma, V., Shukla, A. K., & Manchanda, J. S. (2022). Biofortification—A frontier novel approach to enrich micronutrients in food crops. *Frontiers in Nutrition, 9*, 823408. https://doi.org/10.3389/fnut.2022.823408

Dobrynin, D., Fridman, G., Friedman, G., & Fridman, A. (2021). *Physical mechanisms of plasma-assisted agriculture and food processing.* **Plasma Chemistry and Plasma Processing, 41(4), 735–756.**

Gao, Y., Francis, K., & Zhang, X. (2022). Formation of cold plasma-activated water and its applications in food and agriculture: A review. *Food Research International, 157*, 111246. https://doi.org/10.1016/j.foodres.2022.111246

Garg, M., Singh, V., Sood, S., & Kaur, H. (2023). Biofortification: A multidisciplinary approach to eradicate micronutrient malnutrition. *Frontiers in Plant Science, 14*, 1189452. https://doi.org/10.3389/fpls.2023.1189452

Ghimire, B., Oh, J., Lee, H. Y., & Park, G. (2022). Cold plasma technology for sustainable agriculture: Recent developments and future perspectives. *Applied Sciences, 12*(3), 1125. https://doi.org/10.3390/app12031125

Gond, R., Kumar, A., & Sinha, R. (2023). *Experimental design and optimization for plasma-assisted agriculture.* Agricultural Systems, 206, 103613.

Guo, Z., et al. (2020). *Elemental analysis of biofortified grains using atomic absorption spectroscopy.* Analytica Chimica Acta, 1112, 56–64.

Hensel, O., et al. (2022). Effects of plasma-activated water on soil microbial activity and plant growth. *Journal of Applied Microbiology, 133*(4), 2153–2163. https://doi.org/10.1111/jam.15637

Hu, X., Zhang, Y., Li, J., & Liu, Q. (2023). Plasma-treated irrigation water enhances wheat seed germination and root vigor. Agricultural Water Management, 281, 108232.

Huo, J., Zhou, Q., & Yin, X. (2021). Interactions of plasma-generated reactive species with transition-metal ions in aqueous systems. *Environmental Chemistry Letters, 19*(5), 3813–3825. https://doi.org/10.1007/s10311-021-01208-3

Hussain, A., et al. (2022). *Comparative nutrient physiology of wheat and chickpea under soil micronutrient deficiency.* Field Crops Research, 288, 108706.

Jiang, J., Lu, N., & Li, L. (2021). Reactive oxygen species signaling in plasma-stimulated seed germination. Plant Physiology and Biochemistry, 163, 160–170.

Konchekov, E. M., Yurchenko, A. I., & Filippov, E. A. (2023). Advancements in plasma agriculture: A review of recent applications. *International Journal of Molecular Sciences, 24*(20), 15093. https://doi.org/10.3390/ijms242015093

Kumar, P., Sharma, V., & Singh, A. (2023). Plasma-based micronutrient enhancement for soil fertility improvement. Environmental Technology & Innovation, 32, 103028.

Lindsay, W. L., & Norvell, W. A. (1978). *Development of a DTPA soil test for zinc, iron, manganese, and copper.* Soil Science Society of America Journal, 42(3), 421–428.

Los, A., Shimizu, T., & Steves, D. (2022). Effect of plasma-activated irrigation water on plant productivity. Scientific Reports, 12(1), 2418.

Machala, Z., Janda, M., & Hensel, K. (2021). Interaction of non-thermal plasma with soil microbiota. Frontiers in Agronomy, 3, 674512.

MubarakAli, D., Thirumdas, R., & Deshmukh, R. (2020). Nonthermal plasma in agriculture: Mechanistic insights and applications. Trends in Biotechnology, 38(6), 669–681.

Naik, B., Kumar, S., & Singh, R. (2024). Micronutrient biofortification: A solution for addressing hidden hunger in India. *Sustainable Agriculture Reports,* 5(2), 121–133. https://doi.org/10.1016/j.suar.2024.02.005

Niemira, B. A., et al. (2020). *Nonthermal plasma applications in food and agriculture: Mechanisms and perspectives.* Frontiers in Sustainable Food Systems, 4, 63.

Nishioka, T., et al. (2020). Oxidation of ferrous ions by plasma-generated reactive species in aqueous solution. *Plasma Chemistry and Plasma Processing,* 40(4), 1081–1093. https://doi.org/10.1007/s11090-020-10115-y

Oh, J.-S., Lee, H., Kim, S. B., & Park, G. (2023). Mechanisms of reactive nitrogen species generation in air plasma-activated water. *Chemical Engineering Journal,* 453, 139839. https://doi.org/10.1016/j.cej.2022.139839

Reddy, P., Singh, S., & Kumar, D. (2021). Ion mobility and redox behavior in plasma-activated nutrient solutions. Journal of Electrostatics, 112, 103605.

Rosen, C. J., et al. (2021). *Enzyme-based soil health indicators for sustainable farming.* Soil Biology & Biochemistry, 159, 108304.

Sheera, A., Kaur, J., & Sharma, P. (2025). Biofortification strategies for wheat: Enhancing zinc and iron through integrated nutrient management. *Journal of Cereal Science,* 120, 103612. https://doi.org/10.1016/j.jcs.2025.103612

Singh, A., Verma, R., & Pandey, R. (2024). Nitrate and redox-mediated nutrient cycling under plasma irrigation. Soil and Tillage Research, 247, 106854.

Sood, S., Garg, M., & Singh, V. (2023). Biofortification: An approach to eradicate micronutrient deficiencies. *Frontiers in Plant Science,* 14, 113456. https://doi.org/10.3389/fpls.2023.113456

Stangoulis, J. C. R., & Knez, M. (2022). Biofortification of major crop plants with iron and zinc: Achievements and future directions. *Plant and Soil,* 474(1–2), 57–76. https://doi.org/10.1007/s11104-021-05155-9

Starek, M., et al. (2021). *Plasma chemistry interactions with dissolved metal ions.* Plasma Processes and Polymers, 18(5), 2000213.

Thirumdas, R., Misra, N. N., & Deshmukh, R. R. (2022). Advances in plasma-based food and agriculture applications. Comprehensive Reviews in Food Science and Food Safety, 21(3), 2000–2023.

Traylor, M. J., et al. (2021). *Physicochemical characterization of plasma-activated solutions for agricultural use.* Plasma Chemistry and Plasma Processing, 41, 527–542.

White, P. J., & Broadley, M. R. (2009). Biofortification of crops with seven mineral elements often lacking in human diets. *New Phytologist,* 182(1), 49–84. https://doi.org/10.1111/j.1469-8137.2008.02738.x

Zhang, H., et al. (2023). *Reactive oxygen and nitrogen species in PAW and their biological implications.* Environmental Science & Technology, 57(7), 3125–3134.

Zhang, W., Li, X., & Hu, J. (2023). Electrochemical and physicochemical changes in plasma-treated micronutrient solutions. Chemosphere, 332, 138928.